\documentclass[aps,pre,showpacs,preprint,superscriptaddress]{revtex4}
\usepackage[latin1]{inputenc}
\usepackage[dvips]{graphicx}
\usepackage{amsmath,amssymb,amsfonts,psfrag}

\begin{document}

\title{Bulk viscosity of the Lennard-Jones system at the triple point by dynamical Non Equilibrium Molecular Dynamics}

\author{Pier Luca Palla}
\email{pierluca.palla@dsf.unica.it}
\affiliation{Department of Physics, University of Cagliari, Cittadella Universitaria, I-09042 Monserrato (Ca), Italy and Sardinian Laboratory for Computational Materials Science (SLACS, INFM-CNR)}

\author{Carlo Pierleoni}
\email{Carlo.Pierleoni@aquila.infn.it}
\affiliation{CNISM and Dipartimento di Fisica, Universit\`a de L'Aquila, I-67100 L'Aquila, Italy}

\author{Giovanni Ciccotti}
\email{giovanni.ciccotti@roma1.infn.it}
\affiliation{CNISM and Dipartimento di Fisica, Universit\`{a} di Roma
``La Sapienza'',  Centro Interdisciplinare Linceo "B. Segre", Accademia dei Lincei, Roma, Italy}

\date{\today}
\begin{abstract}
Non-equilibrium Molecular Dynamics (NEMD) calculations of the bulk viscosity of the triple point Lennard-Jones fluid are performed with the aim of investigating the origin of the observed disagreement between Green-Kubo estimates and previous NEMD data. We show that a careful application of the Doll's perturbation field, the dynamical NEMD method, the instantaneous form of the perturbation and the "subtraction technique" provides a NEMD estimate of the bulk viscosity at zero field in full agreement with the value obtained by the Green-Kubo formula. As previously reported for the shear viscosity, we find that the bulk viscosity exhibits a large linear regime with the field intensity which confirms the Lennard-Jones fluid as a genuine Newtonian fluid even at triple point. 
\end{abstract}

\maketitle

\section{Introduction}\label{Intro}
The calculation of the hydrodynamics transport coefficients for model systems is a noticeable success of Molecular Dynamics (MD) \cite{CiccottiHoover}. The standard method to compute linear transport coefficients by Molecular Dynamics simulations makes use of the Green-Kubo formulas \cite{Green,Kubo,HansenMcDonald}. Based on the dissipation-fluctuation theorem, Green-Kubo formulas allow to compute linear transport coefficients from dynamical fluctuations of suitably defined microscopic currents at equilibrium. The Green-Kubo methodology can be easily implemented in a simulation of the equilibrium state and all transport coefficients can be obtained in the same calculations.  

\begin{figure}[htbp]
\begin{center}
\psfrag{meyer}{\tiny{Ref.\cite{M'05} EMD}}
\psfrag{hoover}{\tiny{Ref.\cite{Hoover80_2} NEMD}}
\psfrag{heyes}{\tiny{Ref.\cite{He'84} NEMD}}
\psfrag{lvk}{\tiny{Ref.\cite{LVK'73} EMD}}
\psfrag{lv}{\tiny{Ref.\cite{LV'87} EMD}}
\psfrag{sh}{\tiny{Ref.\cite{SH'85} EMD}}
\psfrag{hohe}{\tiny{Ref.\cite{Hohe'87} EMD}}
\psfrag{hvs}{\tiny{Ref.\cite{HVS'87} EMD}}

\includegraphics[width= 8 cm]{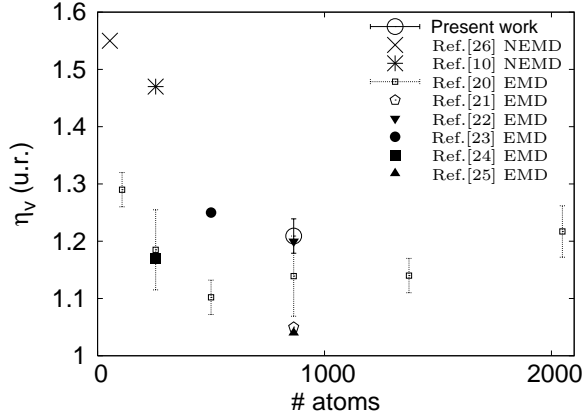}
\caption{Review of previous results for the bulk viscosity of a simple Lennard-Jones fluid close to the triple point. The reported values are plotted as a function of the number of particles in the computations.}
\label{graf_eta_N}
\end{center}
\end{figure}

An alternative approach to the computation of transport coefficients is to mimic the appropriate non equilibrium state. This can be generally obtained by applying a suitable external force and measuring the response related to the corresponding transport coefficient.  Non Equilibrium Molecular Dynamics (NEMD) has been developed along these lines already in the early eighties. It was soon realized that some paradigms of Equilibrium Molecular Dynamics (EMD) had to be relaxed to mimic non equilibrium processes. In particular the use of Periodic Boundary Conditions (PBC), a key ingredient of EMD to minimize finite size effects, is often incompatible with the non equilibrium state of interest. In many interesting cases the external field acts through the boundaries, for instance a thermal gradient or a velocity gradient, and the simulation of such system can require abandoning the use of PBC in favor of less convenient boundaries. This was for instance the case of a systems under the action of a thermal gradient \cite{Trozzi,Gallico} or in a Poiseuille flow \cite{Koplik88}. Non-periodic boundaries however require quite larger systems which had limited the early use of direct non equilibrium methods. To circumvent these limitations the so called "synthetic" NEMD algorithms have been developed and extensively used in the exploration of non equilibrium phenomena\cite{EvansMorris}. 
The general idea behind this class of algorithms is to replace the external force by an effective, PBC compatible, bulk field which, in the limit of vanishing intensity, excites the same response as the original external force. In this way the linear regime can, in principle, be explored without abandoning the use of PBC and therefore avoiding large finite size effects. In the case of fluid flows this technique requires the use of periodic but moving boundary conditions\cite{LeesEdwards,EvansMorris}. It should be noticed that, after restoring the use of PBC, the heat produced by the external bulk field must be removed by a "bulk" thermostatting mechanism such as for instance a Nose-Hoover thermostat. We want to emphasize that the theoretical foundation of this class of algorithm is the Linear Response Theory and their use beyond the linear regime is somewhat arbitrary. In this context the "subtraction technique" \cite{CiccottiJacucciPRIMO,CiccottiJacucciTERZO} is a very useful tool (at least for simple systems in which the response time is not longer than the typical Lyapunov time) to perform the vanishing perturbation limit. %without encountering an exponential grow of the noise to signal ratio. This has been a unique tool to demonstrate the validity of LRT and of the synthetic algorithms.
For almost all transport coefficients a good agreement between GK method and synthetic NEMD methods has been found \cite{RyckaertCiccotti89, MassobrioCiccotti84, HooverCiccottiPaoliniMassobrio85, PaoliniCiccotti87, PierleoniCiccottiBernu87, PierleoniCiccotti90, HonkonnouPierleoniRyckaert92}. The bulk viscosity makes a noticeable exception. In Fig. \ref{graf_eta_N} we show the results of several computations of the bulk viscosity of a Lennard-Jones fluid close to the triple point. Most of these works adopted the Green-Kubo method \cite{LVK'73,SH'85,Hohe'87,HVS'87,LV'87,M'05} and found very similar results. Only two NEMD calculations have been performed so far \cite{Hoover80_2,He'84} and they both provides values of the bulk viscosity 30\%-50\% higher than the Green-Kubo values. Note that finite size effects cannot explain the observed discrepancies.

In the present work we reconsider the calculation of the bulk viscosity of the Lennard-Jones fluid close to the triple point and show that a careful application of the well know Doll's synthetic algorithm provides estimates of the bulk viscosity in full agreement with the Green-Kubo values.
The paper is organized as follows. In section \ref{sec:theory} we provide the necessary theoretical background by discussing both the Green-Kubo formula for the bulk viscosity coefficient and the dynamical NEMD approach we have adopted. Section \ref{sec:NEMD} deals with details of the simulation such as the time dependence of the perturbation, the simulation box and the implementation of the dynamical approach to NEMD. In section \ref{sec:results} we collect our results and in section \ref{sec:conclusions} some concluding remarks. In the Appendix we show that, in the linear regime, the synthetic perturbation for the viscosity is, as generally assumed, proportional to the velocity field produced.

%The main goal of the present work is to show that an accurate application of the NEMD provides estimates of all the transport coefficients in total agreement with the standard Green-Kubo method. In order to obtain such a result we have applied a robust implementation of the NEMD idea, the so-called ``dynamical approach to the non equilibrium molecular dynamics'' \cite{CiccottiJacucciPRIMO,CiccottiJacucciSECONDO,CiccottiJacucciTERZO}. 

%\cite{LVK'73,SH'85,Hohe'87,HVS'87,LV'87,M'05}

\section{Theoretical Background}\label{sec:theory}

\subsection{Hydrodynamics and microscopic identification of local fields}\label{subsec:Hydro-Ident}

The bulk viscosity is one of the transport coefficients introduced in hydrodynamics\cite{HansenMcDonald}. In this theory, the fluid is described by classical fields which, in the case of a simple neutral system, are the mass, momentum and energy density fields. Partial differential equations derived from the continuity equation, supplemented by the so called ``constitutive relations'' and by the local equilibrium hypothesis, provide a closed theoretical framework for the evolution of these fields. The ``constitutive relations" are linear relations between the external forces acting on the system and the excited flows, the coefficients being the transport coefficients specific for each material. The viscosity coefficients, namely the bulk viscosity, $\eta_v$, and  the shear viscosity, $\eta_s$, are defined by the Newton constitutive law
\begin{equation}
\underline{\underline{P}}(\vec r,t)  =  [ \bar p(\vec r,t) - (\eta_v-\frac{2}{3}\eta_s )\vec{\nabla}\cdot \vec v(\vec r,t)]\underline{\underline{I}} -\eta_s [ \vec{\nabla}\vec v(\vec r,t) +  (\vec{\nabla}\vec v(\vec r,t))^{\dag} ]
\label{eq:newton}
\end{equation}
where $\underline{\underline{P}}$ is the pressure tensor, $\vec v$ the velocity field and $\underline{\underline{I}}$ the identity tensor.  $\bar p$ represents the hydrostatic pressure, which, according to the local equilibrium hypothesis, can be expressed  in terms of the mass density, $m(\vec r,t)=m n(\vec r,t)$, and the energy density, $e(\vec r,t)$, by the equilibrium equation of state
\begin{equation}
\bar p(\vec r,t)=P_{eq}(mn(\vec r,t),e(\vec r,t))
\label{eq:EOS}
\end{equation}
where $n(\vec r,t)$ is the local number density.
If the velocity field reduces to the particular form $\vec v(\vec r ,t ) =  \vec r f(t)$, with $f(t)$ a yet unspecified function of time, the Newton law reduces to
\begin{equation}
\frac{1}{3}Tr[\underline{\underline{P}} - \bar p\underline{\underline{I}}] = - 3 \eta_v f(t)\label{macro_eta}
\end{equation}
where the symbol $Tr$ stands for the trace of the tensor.

According to Irving and Kirkwood \cite{Kirk}, any macroscopic hydrodynamic field $J(\vec r ,t )$ can be obtained from the statistical average, over the time dependent ensemble, $\rho(\Gamma,t)$, of a microscopic observable $\hat J(\Gamma)$, where $\Gamma=\left\{\vec r_i, \vec p_i \right\} (i=1,N)$ is the phase-space point of the N particles system. In the specific case of the pressure tensor we have\cite{HansenMcDonald}
\begin{equation}
\underline{\underline{P}}(\vec r,t)  =  \left< \sum_{i=1}^N\delta(\vec r-\vec r_i)\left[\frac{\vec p_i \vec p_i}{m} + \vec r_i \vec F_i \right] \Big| \rho(\Gamma,t) \right>
\label{eq:pressure_tensor}
\end{equation}
where $\vec F_i$ is the internal force acting on particle $i$.
The number and energy densities can also be expressed as 
\begin{eqnarray}
n(\vec r,t) &=& \left< \sum_{i=1}^N\delta(\vec r-\vec r_i) \Big| \rho(\Gamma,t) \right> \\
e(\vec r,t) &=& \left< \sum_{i=1}^N\delta(\vec r-\vec r_i) \left[\frac{|\vec p_i| ^2}{2m} +\frac{1}{2}\sum_{j\neq i}\phi(r_{ij}) \right ]  \Big| \rho(\Gamma,t) \right>
\end{eqnarray}
where $\phi(r)$ represents the pair potential of our model system.
The local hydrostatic pressure $\bar p(\vec r,t)$ is the trace of the pressure tensor and, within the local equilibrium hypothesis, its fluctuations around the equilibrium value $p_0(n_0, e_0)$ can be expressed as
\begin{equation}
\bar p(\vec r,t) = p_0 +\frac{\partial P_{eq}}{\partial n}\Big |_{n_0} \left( e(\vec r,t) -  e_0\right) + \frac{\partial  P_{eq} }{\partial n }\Big |_{e_0}  \left( n(\vec r,t) -  n_0 \right)
\label{bar_p_mac}
\end{equation}
The bulk viscosity coefficient in terms of statistical averages is then obtained, for velocity fields of the form prescribed above, by replacing the fields in Eq.(\ref{macro_eta}) with the appropriate dynamical ensemble averages. Furthermore, in order to ensure the validity of local equilibrium, it is required to take the ``hydrodynamic limit'':
\begin{equation}
\eta_v=-\lim_{\omega \rightarrow 0}\lim_{k \rightarrow 0}
\frac{ Tr[\underline{\underline{ \tilde{P}}}(\vec{k},\omega) -  \tilde{\bar p}(\vec{k},\omega)\underline{\underline{I}}]}
{ 9\tilde{f}(\omega)}
\label{eta-mic-mac}
\end{equation}
where $\tilde{J}(\vec k, \omega)= \frac{1}{V}\int \int d\vec r \;dt \exp(i \vec k \cdot \vec r )\exp(i\omega t) J(\vec r,t )$ is the usual Fourier transform (in space and in time) of the field.

\subsection{Green-Kubo formula}\label{subsec:GK}
Equation (\ref{eta-mic-mac}) expresses the bulk viscosity as an average on the non equilibrium distribution $\rho(\Gamma,t)$. When the system is close to equilibrium (linear hydrodynamics) it is possible to rewrite it by the Green-Kubo formula\cite{Mori,McQuarry}:
\begin{equation}
\eta_v= \beta V \int_{0}^{+\infty}dt < \hat J(t)\hat J(0)  >_0 \label{eq:greenkubo}
\end{equation}
with
\begin{widetext}
\begin{equation}  
\hat{J}(t) = (\hat{\mathcal {P}}(t)  - <\hat{\mathcal {P}}>  ) - \frac{1}{V} \left\{ \frac{\partial P_{eq}}{\partial e}\Big |_{n} ( \hat{\mathcal H}(t) -  <\hat{\mathcal H}>) 
%+  \frac{\partial P_{eq} }{\partial n }\Big |_{u}( \hat{\mathcal N}(t) - < \hat{\mathcal N} >) 
\right\}  
\label{eq:flux}
\end{equation}
\end{widetext}
where $V, e$ are the volume and internal energy per unit volume of the system, respectively. 
%$\hat{\mathcal N}$ is the instantaneous number of particles while 
$\hat{\mathcal {P}}$ and $\hat{\mathcal H}$ are the dynamical variables corresponding to the thermodynamic pressure and energy
\begin{eqnarray}
\hat{\mathcal {P}}(t) & = & \lim_{k \rightarrow 0} \frac{1}{3V} Tr (\tilde{\hat{\underline{\underline P}}}(\vec{k},t) )\label{DefmatP}=\frac{1}{3V}\sum_{i=1}^N Tr\left[\frac{\vec p_i \vec p_i}{m} + \vec r_i \vec F_i \right] \label{eq:pressure} \\
\hat{\mathcal H}(t) & = & \lim_{k\rightarrow 0}\tilde{\hat e}(\vec{k},t)\label{DefMatH}=\sum_{i=1}^N \left[\frac{|\vec p_i| ^2}{2m} +\frac{1}{2}\sum_{j\neq i}\phi(r_{ij}) \right ]
\end{eqnarray}
Eq. ($\ref{eq:flux}$) is the general expression of the current related to the bulk viscosity coefficients in the general case in which the energy fluctuates. If experiments are conducted in the microcanonic ensemble  the energy fluctuations vanishes and the more familiar expression of the Green-Kubo formula is obtained\cite{HansenMcDonald}.

\subsection{``Doll's" perturbation and the `dynamical approach' to Non Equilibrium Molecular Dynamics}\label{subsec:NEMD}
As described in the introduction, the alternative route to transport properties is to consider the system subjected to an external perturbation able to mimic the ``thermodynamic force" which excites the appropriate nonequilibrium flux inside the system. 
In the present case the ``thermodynamic force'' is the macroscopic velocity gradient, $\vec{\nabla} \vec v$, while the corresponding flux is the deviation of the pressure from its local equilibrium value,  $\frac{1}{3}Tr[\underline{\underline{P}}] - \bar p$. The bulk external force to be used in such experiments is known as ``Doll's'' perturbation:
\begin{equation}
\hat H'(\Gamma,t)=\sum_i^N \vec r_i \vec p_i: \left(\vec{\nabla}\vec u(\vec r_i,t)\right)^T \label{dolls} 
\end{equation}
where $\vec{\nabla}\vec u(\vec r,t)$ is the required external field. This perturbation was proposed for the first time by Luttinger \cite{Luttinger} and adopted in a Molecular Dynamics simulation by Hoover et al.\cite{Hoover80_1,Hoover80_2}. In the Appendix we will show that, in the linear regime, the macroscopic velocity field $\vec v(\vec r,t)$ induced by this perturbation coincides with imposed external constraint $\vec u(\vec r,t)$ in the long wave-lenght limit
\begin{equation}
\tilde{\vec{v}}(\vec k=0,\omega)=\tilde{\vec{u}}(\vec k=0,\omega) \label{vVSu}
\end{equation}
Once we are able to induce the required hydrodynamic flux, we need a procedure to compute the average of the response on the non equilibrium ensemble. 

To this aim we can exploit the ``Onsager-Kubo'' relation\cite{CiccottiJacucciPRIMO,CiccottiJacucciTERZO,CiccottiPierleoniRyckaert90}. Calling $S(t)$ the time evolution operator of the perturbed dynamics, the following relation holds for the non equilibrium average of the generic microscopic flux $\hat{J}$
\begin{eqnarray} 
 < \hat J>_t  &\equiv&  \int \hat J (\Gamma) \rho(\Gamma,t) d\Gamma  = \int \hat J (\Gamma)  S^{\dagger}(t)\rho_0(\Gamma) d\Gamma= \nonumber \\ 
 & = & \int S(t) J (\Gamma)\rho_0(\Gamma) d\Gamma  = \int \hat J (t)\rho_0(\Gamma) d\Gamma \equiv <\hat J(t)>_0 
\label{Ons-kubo}
\end{eqnarray}
where $S^{\dagger}(t)$ is the adjoint of $S(t)$ and $\rho_0(\Gamma)$ is the ensemble distribution at the time $t=0$. If the perturbation is switched on at time $t=0$ from an equilibrium state, $\rho_0$ is the equilibrium distribution and the Onsager-Kubo relation (\ref{Ons-kubo}) allows us to compute the required average in a rigorous way. Indeed, via standard equilibrium MD simulation, we can obtain a set of statistically independent configurations $\{\Gamma_i \}$ distributed according to $\rho_0(\Gamma)$. Starting from those configurations we can follow the evolution of the system under the perturbed dynamics and obtain the required nonequilibrium average as the average of the evolved observable over the initial distribution according to the Onsager-Kubo relation.

When a large perturbation is applied a thermostatting mechanism needs to be added to the equation of motion and the response can depend on it. Conversely for vanishingly small perturbations the standard form of linear response theory \cite{Kubo,HansenMcDonald} holds and the response depends only on the applied perturbation. In this limit however an additional numerical problem is encountered. The fluctuations of the microscopic variables are quite large and dominate the response in the limit of vanishing perturbations. In simple systems, where the response time is comparable to the Lyapunov time of the exponential divergence of nearby starting trajectories,  the ``subtraction technique" can be used to extract the signal out of the statistical noise \cite{CiccottiJacucciPRIMO,CiccottiJacucciTERZO}. If $S_0(t)$ is the evolution operator representing the unperturbed dynamics with $<S_0(t)\hat J>_0=0$, the average values $<S(t)\hat J-S_0(t)\hat J >_0$ and $<S(t)\hat J>_0$ are equal. On the other hand, their variances are very different: the thermal fluctuations of $S(t)\hat J$ and $S_0(t)\hat J$ are highly correlated and therefore largely cancel each others. This is true for times short enough. At large times the variance of the difference estimator becomes twice the variance of the simple estimator.

\section{Simulation Technique}\label{sec:NEMD}

%In this section we discuss the choice of the time dependence of the external perturbation and some technical aspects of the NEMD computations such as the generalization of the standard Periodic Boundary Conditions. Moreover, some features of the NEMD computation specific for the bulk viscosity coefficient are emphasized.

\subsection{Impulsive external field}\label{subsec:impulsive}

As mentioned above, the current associated to the bulk viscosity to be used both in the Green-Kubo formula or in the NEMD experiments depends on the statistical ensemble chosen to conduct the experiment. In all previous equilibrium calculations, as well as in the present one, the microcanonical ensemble was chosen since the current reduces to the pressure tensor fluctuations without the need of evaluating the additional term related to the energy fluctuations. 

In NEMD, the Doll's perturbation for the bulk viscosity is a pure contraction or expansion of the volume so that a constant perturbation in time, $f(t)=\epsilon$,  will correspond to an exponential contraction/expansion of the volume (see next subsection), obviously an impractical way to extract the bulk viscosity coefficient. Alternatives forms of $f(t)$ can be an oscillating function $f(t)=\epsilon \sin(t)$ and an impulsive perturbation $f(t)=\epsilon\delta(t)$, where $\epsilon$ is the intensity of the field. The oscillating form, used in previous NEMD calculations, requires thermostatting the nonequilibrium trajectory in order to reach a steady state and the extended form of the current with the energy fluctuation term must be used. On the other hand, the impulsive perturbation acts on the system for an infinitesimal time, no thermostatting mechanism is necessary, and the dynamics after the impulse is the equilibrium dynamics for the isolated system. This fact greatly simplifies the NEMD experiment and the form of the flux, eq.(\ref{eq:flux}).       
For each initial configuration $\Gamma_i$, the energy and the volume of the system change from their initial values ${\mathcal H}_0$ and $ V_0$ of the equilibrium system, to the time independent values ${\mathcal H}'_i=\hat{\mathcal H}_i(t>0^+)$ and $ V'=\hat V(t>0^+)$ (note that all replicas undergoes the same volume change). In order to obtain the appropriate flux in Eq. (\ref{eq:flux}) we have to calculate only the dynamical variable $\hat{\mathcal P}(t)$ and its average asymptotic value:
\begin{equation}
 p_{\infty} = \lim_{t\rightarrow\infty}<\hat{\mathcal P}>_t %=P(<{\mathcal H}'>_0,<n'>_0) \label{bar_p_impulso}
\end{equation}
where $<\dots>_t$ are averages over the non-equilibrium distribution at time $t$. Note that $p_{\infty}$ is not a properly defined pressure because the corresponding ensemble is not well defined since each member of the ensemble has a different energy. It is rather the asymptotic large time value of $<\hat{\mathcal P}>_t$ that needs to be subtracted to it in order to make the current integrable in time to provide the associated transport coefficient.

\subsection{Periodic Boundary Conditions}\label{subsec:PBC}
The explicit form of the Doll's perturbation field with a homogeneous velocity gradient 
$\nabla u= f(t) \mathbb{I}$, is not compatible with the periodic boundary conditions of MD. However the periodicity of the system can be restored if we allow the box matrix ${\underline{\underline H}}$ to evolve according to the external flow\cite{ryckaert90Aachen}
\begin{equation}
\dot{\underline{\underline H}}(t) = \underline{\underline{\nabla u}} \cdot \underline{\underline H}(t) \label{eqH} 
\end{equation}
Starting with a cubic cell of edge $L_{\alpha}(0)=L_0 (\{\alpha=x,y,z\})$, and applying a velocity field $\nabla u = \epsilon \delta(t) \mathbb{I}$ we get
\begin{equation}
L_{\alpha}(t \geq 0^+)=L_0e^{\epsilon} \simeq L_0(1+\epsilon)
\end{equation}
Similarly, the perturbation induces a discontinuity in the trajectory of the system
\begin{eqnarray}
\dot{\vec{r_i}} & = &  \frac{\vec{p_i}}{m} + \epsilon  \delta(t) \vec{r_i} \nonumber \\
\dot{\vec{p_i}} & = &  \vec{F_i} -\epsilon \delta(t) \vec{p_i}  \label{eqmoto}
\end{eqnarray}
which correspond to
\begin{eqnarray}
\vec{r_i}(0^+)  & = &\vec{r}_i(0^-)e^{\epsilon}  \simeq  \vec{r}_i(0^-)(1 + \epsilon)  \nonumber\\
\vec{p_i}(0^+)  & = &\vec{p}_i(0^-)e^{-\epsilon}  \simeq \vec{p}_i(0^-)(1 - \epsilon) \label{eq0-0+}
\end{eqnarray}
Therefore, the effect of the impulsive perturbation is to apply a homogeneous contraction (expansion) of the position space and an homogenous expansion (contraction) in the momentum space of the system.
Substituting eqs ($\ref{eq0-0+}$) in the perturbed Hamiltonian $\hat{\mathcal H}(0^+)=\hat{\mathcal H}_0+\epsilon\sum_i\vec{r}_i(0^+)\cdot\vec{p}_i(0^+)$ and using eq. (\ref{eq:pressure}), the variation of energy induced by the impulsive field is
$$
\hat{\mathcal H}(0^+)-\hat{\mathcal H}(0^-)= -\hat{\mathcal P}(0^-) dV+ {\mathcal O}(\epsilon^2)
$$
where $V$ is the volume of the system.
Taking the ensemble average over the equilibrium distribution at $t=0^-$, we recover the first principle of thermodynamics 
\begin{equation}
\Delta E=-pdV
\label{primoprincipio}
\end{equation}

\subsection{Bulk viscosity computation}\label{subsec:bukviscosity}
 
With the impulsive field of previous section, eq. (\ref{eta-mic-mac}) for the bulk viscosity reduces to
\begin{eqnarray}
\eta_v & = &\lim_{\omega \rightarrow 0}\lim_{k \rightarrow 0}-\frac{  \frac{1}{3}Tr[\underline{\underline{ \tilde{P}}} -  \tilde{\bar p}\underline{\underline{I}}]}{Tr(   \tilde{\vec{\nabla}}\vec{v}    )  } = \nonumber\\
& = &\lim_{\epsilon \rightarrow 0} - \frac{1}{3 \epsilon}\int_0^{\infty}\left[< \hat{\mathcal {P}} >_t  -  p_{\infty}\right] dt 
\label{quasi-last-eta}
\end{eqnarray}
Using the Onsager-Kubo relation to rewrite nonequilibrium ensemble averages in terms of equilibrium averages of evolved observables, and applying the subtraction technique described above, we obtain 
\begin{equation}
\eta_v =\lim_{\epsilon \rightarrow 0} - \frac{1}{3\epsilon}\int_0^{\infty}\left\{\left[< S(t)\hat{\mathcal {P}}-  S_0(t)\hat{\mathcal {P}}>_0\right] - \left[p_{\infty}-p_0\right] \right\}dt \label{eta_last}
\end{equation}
where $S(t)$ and $S_0(t)$ are the evolution operators for the perturbed and equilibrium dynamics respectively. Note that $p_0=P_{eq}(\hat{\mathcal{H}}_0,n_0)$ is the pressure of the equilibrium microcanonical system, while as mentioned above, $p_{\infty}$ is not a properly defined averaged pressure.

\subsection{Simulations scheme and numerical details}\label{subsec:numerics}

As already mentioned, the operative procedure to compute eq. (\ref{eta_last}) is to select a set of statistically uncorrelated equilibrium configurations of the system $\{\Gamma_i\}$, for each member of the set to generate both the equilibrium and the non equilibrium evolutions, and to compute the term $[< S(t)\hat{\mathcal {P}}-  S_0(t)\hat{\mathcal {P}}>_0]$ up to a time $t_{r}$, as the arithmetic mean over the set. The time $t_{r}$ is the typical time the individual non equilibrium system $C_i$ of energy $\hat{\mathcal{H}}_i(0^+)$  takes to relax from the perturbed initial configuration to its equilibrium state. The additional offset term, involving $p_{\infty}$, is the stationary value of that average beyond $t_{r}$. In order to reduce the statistical noise on this term we have computed, for each individual nonequilibrium trajectory, the time average of the microscopic ``pressure'' between $t_{r}$ and $2~t_{r}$ and we have estimated the offset as the arithmetic mean of those time averages over all nonequilibrium trajectories. 
\begin{figure}
 \includegraphics[width=8cm ]{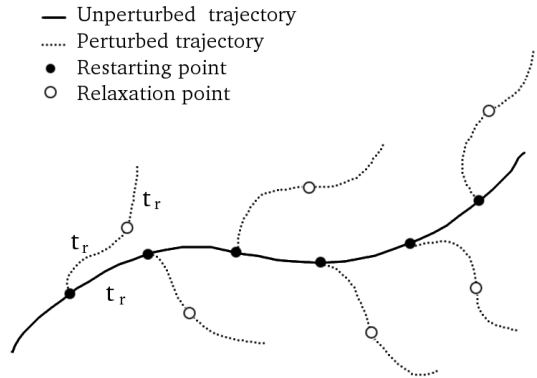}
 \caption{\label{figfish}Simulations scheme: a long unperturbed trajectory of the system is used to sample initial conditions for segment of perturbed trajectories. Each integration of the perturbed equations of motion has been integrated for twice the estimated signal relaxation time. This is needed to evaluate the thermodynamic pressure of the segments.}
 \end{figure}
 
The system considered in this work is a simple fluid of $N=864$ particles interacting by the Lennard-Jones potential, in a thermodynamic state close to the triple point. In the following all quantities will be expressed in Lennard-Jones units: $\epsilon=1, \sigma=1, m=1$. The potential has been truncated at $r_{cut} = 2.5$ and shifted to avoid discontinuity at $r_c$. Moreover for $r\in[2.4:2.6]$ the potential has been replaced with a cubic polynomial in order to avoid discontinuities in the forces at the cutoff. The equations of motion have been integrated through a Velocity Verlet algorithm with an integration step $h=0.004436$. 

The unperturbed trajectory (see figure $\ref{figfish}$) has been integrated for $9.9\times10^6$ integration steps.  No thermalizing device is added to the equilibrium dynamics so that a sampling of the microcanonical ensemble is obtained. This equilibrium trajectory was used to compute the bulk viscosity through the Green-Kubo formula. The time of saturation of the Green-Kubo integral, i.e. the decorrelation time of the pressure fluctuations at equilibrium, has been used as an estimate of the relaxation time $t_{r}$. The set of initial equilibrium configurations are therefore selected as equilibrium configurations a time $t_{r}$ apart from each other along the equilibrium trajectory. In order to calculate the response  of the system to the impulsive external field, $S(t)\hat{\mathcal {P}}$, the perturbed trajectories have been integrated for a time $t_{r}$ and extended to $2t_{r}$ in order to evaluate the offset term (see figure $\ref{figfish}$).

\section{Results}\label{sec:results}

The thermodynamic state of the equilibrium system is reported in table $\ref{TStatetab}$. This state is very close to that used in previous studies\cite{LVK'73,SH'85,Hohe'87,HVS'87,LV'87,M'05,Hoover80_2,He'84}.  
% punto triplo era: ($\rho_t=0.85 u.r.,\; T_t=0.68 u.r.$) diverso da Hoover state
\begin{table}
\caption{\label{TStatetab} Thermodynamic equilibrium state. The simulated system is in a state very close to that used in most of previous MD studies of the bulk viscosity.}
\begin{ruledtabular}
\begin{tabular}{l l l l}
%Density [u.r.]& Energiy/N [u.r.] & Kin.En./N [u.r.]& Pot.En./N [u.r.]& Temp. [u.r.]& Pressure [u.r.]\\
Density (r.u.) & Energy/N (r.u.)  & Temp. (r.u.)& Pressure (r.u.)\\
\hline
%0.8442 & -4.112561(5) &  1.08042(6)  & -5.19298(6)  & 0.72111(4)  & 0.8978(3)\\
0.8442 & -4.112561(5) & 0.72111(4)  & 0.8978(3)\\
\end{tabular}
\end{ruledtabular}
\end{table}

\subsection{Green-Kubo results}\label{subsec:GKresults}

\begin{figure}[htbp]
\begin{center}
\includegraphics[ width=8cm]{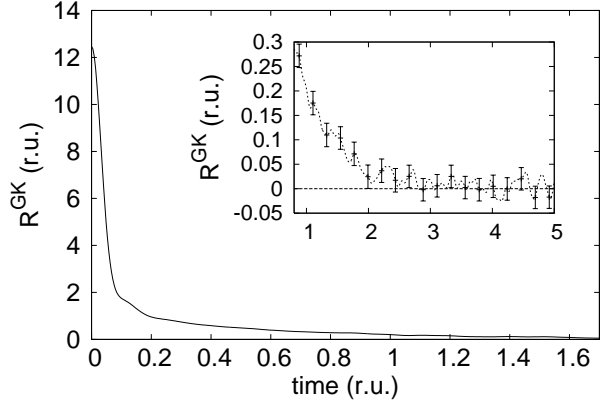}
\caption{ Green-Kubo integrand (see Eq.($\ref{eq:greenkubo}$) ). In the inset we show an enlargement of the tail.}
\label{graf-GKintegrand}
\end{center}
\end{figure}

\begin{figure}[htbp]
\begin{center}
\includegraphics[width=9cm]{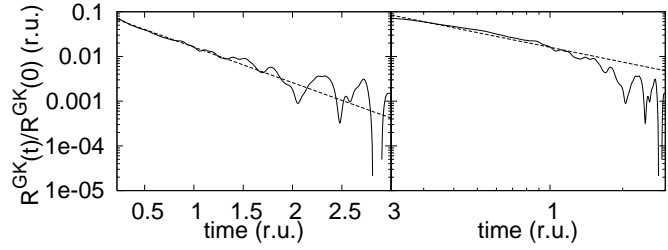}
\caption{Tail of the Green-Kubo integrand $R^{GK}(t)$ and its fitting functions. Data are normalized by the initial value $R^{GK}(0)$. The fitting range is  $t \in [0.22:2 ]$. In the left panel we show (in linear-log scale) the exponential fit, while in the right panel we show (in log-log scale) the power law fit.}
\label{graf-interpolations}
\end{center}
\end{figure}

In Fig. \ref{graf-GKintegrand} we report the Green-Kubo integrand $R^{GK}(t)=\beta V<\hat{\mathcal {P}}(t)\hat{\mathcal {P}}(0)>_0$. After a rapid relaxation in about 0.22 time units, the curve exhibits a long time tail which vanishes only beyond t=2 (see the inset). Although the noise level on the time correlation function is quite small, the noise on its time integral, as obtained by a simple trapezoidal rule, results to be quite large because of the long tail. In order to get a smoothed signal we have attempted to replace the data beyond t=0.22 with two different analytic functions, fitted to the data in the time interval [0.22:2.0]. We have assumed a power law behavior $R^{GK}_v(t)\sim a t^{-b}$ and an exponential behavior $R^{GK}(t)\sim a~e^{-bt}$ (the power law behaviour is compatible with Hoover's hypothesis \cite{Hoover80_2}: $\eta^{GK}_v(\omega)=a' + b'\sqrt{\omega}$, i.e. $R^{GK}(t)\sim t^{-\frac{3}{2}}$). Values of the fitting parameters are reported in table \ref{tab-Interpolazioni1} while the data and the fitting functions are compared in figure \ref{graf-interpolations}.

\begin{table}
\caption{\label{tab-Interpolazioni1} Fitting parameters for the tail of the Green-Kubo integral. }
\begin{ruledtabular}
\begin{tabular}{l l l l l}
           & fitting interval & $a$(r.u.)  &  $b$(r.u.)   & $\frac{\chi^2}{ndf}$ \\
\hline
$\eta=a t^{b}$ & $t\in[0.22:2.0]$& 0.0160(2) &  -1.09(1)& 1e-05    \\
$\eta=a e^{-b t}$ &$t\in[0.22:2.0 ]$& 0.1004(6) & 1.83(1) & 2e-06  \\
\end{tabular}
\end{ruledtabular}
\end{table}

\begin{figure}[htbp]
\begin{center}
\includegraphics[width=8cm]{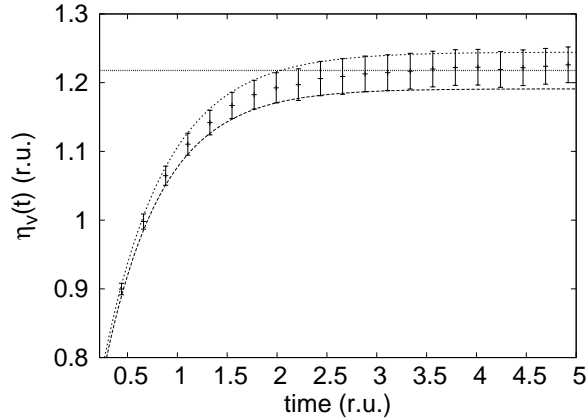}
\caption{Green-Kubo integral: $\eta_v(t)=\int_0^t~ds R^{GK}(s)$. %Exponential fits to the integrand in the range $t \in [0.22:3.55 ]\;(u.r.)$ are performed both for the maximum and the minimum curve. Their integral give rise to the error bars shown in the plot. 
The horizontal plateau value is our Green-Kubo estimate of the bulk viscosity coefficient.}
\label{graf-GKintegral}
\end{center}
\end{figure}
From table \ref{tab-Interpolazioni1} and Figure \ref{graf-interpolations}, we conclude that the exponential behaviour is a better representation of our data. 
% QUI NON CAPISCO NULLA !!!!!
%So the integral of these data up to time $t$ has been fitted through the following form:  $d(1-e^{ct})$. The value of the parameter $d$ is adopted as the estimate of the integral, i.e of the bulk viscosity $\eta_v$. The results of the last interpolation is reported in Fig.$\ref{graf-GKintegral}$ and  the obtained estimate of the balk viscosity coefficient is:
Integration of the correlation function supplemented by our best exponential fit provides the behaviour in figure \ref{graf-GKintegral} and the following value for the viscosity
$$
\eta^{GK}_v= 1.22 \pm 0.03
$$   
%Finally, we can estimate from the inset in Fig. $\ref{graf-GKintegrand}$ the decorrelation time $t_{rel} \simeq 2.5 r.u.$.

\subsection{NEMD results}

\begin{figure}[htbp]
\begin{center}
\includegraphics[width=8cm]{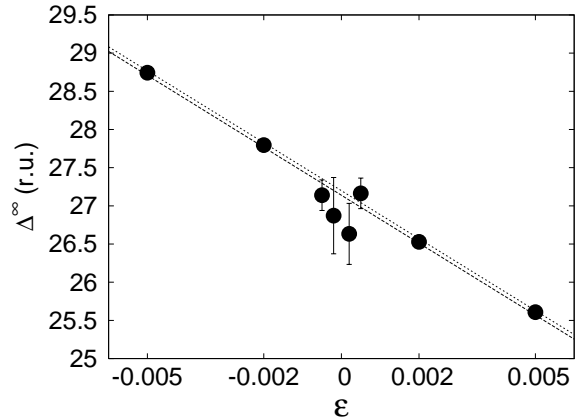}
\caption{Trend of the offset, ${\Delta}^{\infty}$, defined in Eq.($\ref{DeltaInf}$), as a function of the applied deformation $\epsilon$. For $\mid{\epsilon}\mid < 0.002$ the values are affected by a large error. An interpolation of the other data is used to reduce the noise in this region.}
\label{graf-interpOffset}
\end{center}
\end{figure}

In table \ref{tab-segments} we report the details and thermodynamic results from the performed nonequilibrium experiments. Note that positive $\epsilon$ corresponds to nearly adiabatically expanded systems and therefore to reduced temperatures (remember that the system remain isolated after the impulsive external perturbation at t=0). 
In the second last column we report the average variation of the total energy $\Delta E$ due to the impulsive perturbation. Comparing the data in the last two columns we can verify the validity of Eq. ($\ref{primoprincipio}$) in the limit of small $\epsilon$ ($|\epsilon|<0.002$).
\begin{table}
\caption{\label{tab-segments} Thermodynamic properties of the system subjected to the external perturbation. As for the $p_{\infty}$, the values of the temperature $T$ and of the energy variation $\Delta E$ are obtained as the arithmetic mean of the corresponding properties over all non equilibrium trajectories, see Section \ref{subsec:numerics}. By comparing the data in the last two columns, we can verify the validity of the Eq.($\ref{primoprincipio}$). As explained in section \ref{subsec:PBC}, this equation states the consistency of the ``'Doll's'' perturbation with the first principle of thermodynamics.}
\begin{center}
\begin{ruledtabular}
\begin{tabular}{l l l l l l }
$\epsilon$ & $N_{seg}$ & N/V & T & $\Delta E/N$ & $-p_0dV/N$  \\
\hline
0.05  &9000& 0.7292 & 0.5408(2)     &  0.2191(1)   &  0.2023(2)    \\
0.02  &9000& 0.7955 & 0.6174(2)     &  0.01049(6)  &  0.02648(6) \\
0.005 &13000& 0.8317 & 0.6912(1)    &   -0.01080(2) &  -0.02160(1)   \\
0.002 &16400& 0.8392 & 0.70892(9)   &  -0.005538(5) &  -0.006386(5)   \\
5$\times10^{-4}$ &16500& 0.8429 & 0.71726(8)  &  -0.001539(1) &  -0.001511(2)  \\ 
2$\times10^{-4}$ &16500& 0.8437 & 0.71909(9)  &  -6.284(5)$\times10^{-4}$  & -6.263(5)$\times10^{-4}$  \\
 0  &---& 0.8442  & 0.72111(4) &  - & - \\
-2$\times10^{-4}$ &12000 & 0.8447 & 0.7220(1) &  6.432(6)$\times10^{-4}$  &  6.352(5)$\times10^{-4}$    \\
-5$\times10^{-4}$ &16500 & 0.8455 & 0.72335(9) &   0.001645(1)         & 0.001660(1) \\
-0.002  &16400 & 0.8493 & 0.7338(1) &  0.007222(5)          & 0.006360(5) \\
-0.005  &13000 & 0.8570  & 0.7533(1) &   0.02133(2)           & 0.01585(1) \\
-0.02   &9000  & 0.8969  & 0.86865(2) &  0.1603(1)            & 0.06218(7) \\
-0.05   &9000  & 0.9846  & 1.1894(5)  &  0.9154(3)            &  0.6539(5)   \\
\end{tabular}
\end{ruledtabular}
\end{center}
\end{table}
In order to analyze the response of the system, we need to separately discuss the various contribution to the integrand in eq. (\ref{eta_last}). For sake of clarity, let us define the following quantities
\begin{eqnarray}
\Delta(t) &=& \frac{\left<  \left( S(t)\hat{\mathcal{P}} -S_0(t)\hat{\mathcal{P}} \right)\right>_0}{-3\epsilon } \label{Deltat} \\
\Delta^{\infty} &=&  \frac{p_{\infty}-p_0 }{-3\epsilon } \label{DeltaInf}\\
R(t,\epsilon)&=&\Delta(t)-\Delta^{\infty}  \label{R-t-eps}\\
\eta_v(t,\epsilon)&=&\int_0^t ds R(s,\epsilon)\label{Eta-t-eps}
\end{eqnarray}
The newtonian bulk viscosity is the zero field-infinite time limit of $\eta_v(t,\epsilon)$.
In table \ref{tab-finale} and in figure \ref{graf-interpOffset}, we report data for $\Delta(0)$ and $\Delta^{\infty}$ defined in Eqs.(\ref{Deltat}) and (\ref{DeltaInf}).

\begin{figure}[htbp]
\begin{center}
\includegraphics[width= 8 cm]{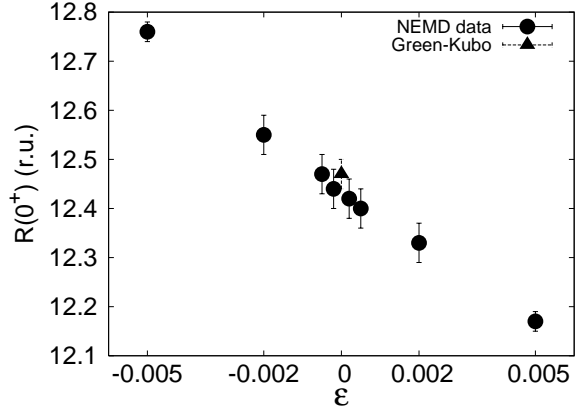}
\caption{Initial time response $R(0^+)$ versus the perturbation. NEMD data tends to the corresponding Green Kubo value in the limit of small $\epsilon$.}
\label{graf-CoEpsilon}
\end{center}
\end{figure}

%\begin{table}
%\caption{\label{tab-delta}\textbf{forse questa tabella potrebbe essere omessa}}
%\begin{ruledtabular}
%\begin{tabular}{l l l l}
%$\epsilon$  & $N_{seg}$ & ${\Delta}(0)$ & ${\Delta}^{\infty}$ \\
%\hline
%0.05   & 9000 &  25.132(5)   & 13.757(6)  \\
%0.02   & 9000 & 32.884(5)   & 21.382(7) \\
%0.005  & 13000 &  37.778(4) & 25.61(2)  \\
%0.002  & 16400 &  38.863(4)   & 26.53(4)  \\ 
%0.0005 & 16500 &  39.415(4)   & 27.1(2)  \\
%0.0002 & 16500 &  39.527(4)   &  26.6(4)  \\
%-0.0002 & 12000 & 39.663(5)  & 26.9(5)  \\
%-0.0005 & 16500 & 39.787(4)   & 27.2(2)  \\
%-0.002  & 16400 & 40.351(4)  & 27.80(4)  \\
%-0.005  & 13000 & 41.499(5) & 28.74(2)  \\
%-0.02   & 9000 & 47.888(6)  & 34.116(7)  \\
%-0.05   & 9000 &  64.387(8) & 47.770(5)   \\
%\end{tabular}
%\end{ruledtabular}
%\end{table}
We note that the error on $\Delta^{\infty}$ grows when $|\epsilon|$ decrease while the error on $\Delta(0)$ remains roughly constant. This is the effect of the subtraction technique that improves the signal to noise ratio for short $t$ only while has no effect at large time where $\Delta^{\infty}$ has to be calculated. A less noisy estimate of $\Delta^{\infty}$ for $|\epsilon| < 0.002$ can be obtained by linear interpolation of the less noisy data at larger absolute values of the perturbation (see Fig. $\ref{graf-interpOffset}$ ).
In Fig. $\ref{graf-CoEpsilon}$ we show the values of the integrand in Eq.($\ref{eta_last}$) at $t=0^+$, i.e. $R(0)= \Delta(0)-\Delta^{\infty}$. At $\epsilon = 0$ we display $R^{GK}(0)$. As predicted by Linear Response Theory, we observe that  the NEMD response tends to the quadratic fluctuations of the pressure at equilibrium in the limit $|\epsilon|\rightarrow 0$. 
\begin{figure}[htbp]
\begin{center}
\includegraphics[width= 8 cm]{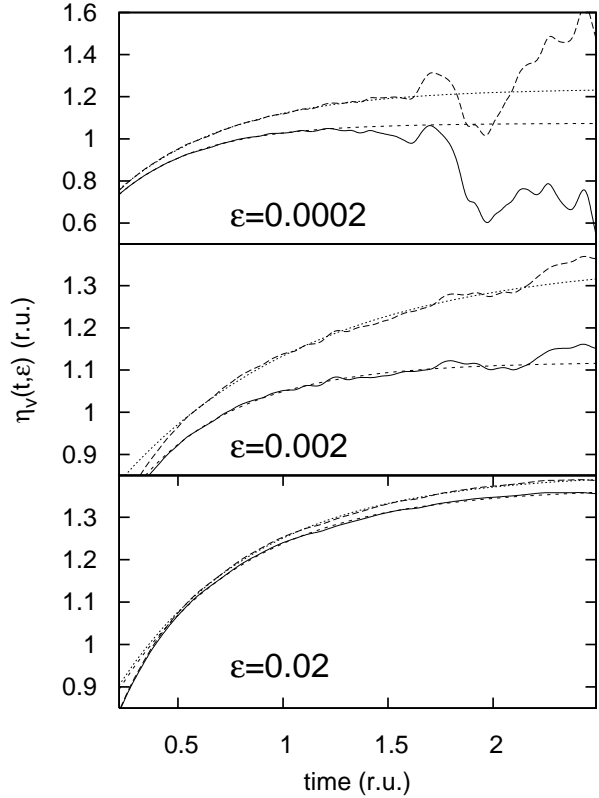}
\caption{Integral $\eta_v(t,\epsilon)$ of the response $R(t,\epsilon)$, see Eqs.\ref{Eta-t-eps},  for some values of $\epsilon$. At large time and for small values of $\mid{\epsilon}\mid$, the subtraction technique is not able to reduce the noise in the signal.}
\label{graf-all-Int}
\end{center}
\end{figure}
In Fig. $\ref{graf-all-Int}$  we show the estimates of the $\eta_v(t,\epsilon)$ curves. We note that, as $\mid{\epsilon}\mid$ decreases, the noise level at large time increases considerably. This signals again the limit of applicability of the subtraction technique. Similar to the Green-Kubo case of previous section, a less noisy estimator of the viscosity is obtained by integrating a response function in which the long time tail is replaced by an exponentially decaying behaviour fitted to the data at large time. 
\begin{figure}[htbp]
\begin{center}
\includegraphics[width= 8 cm]{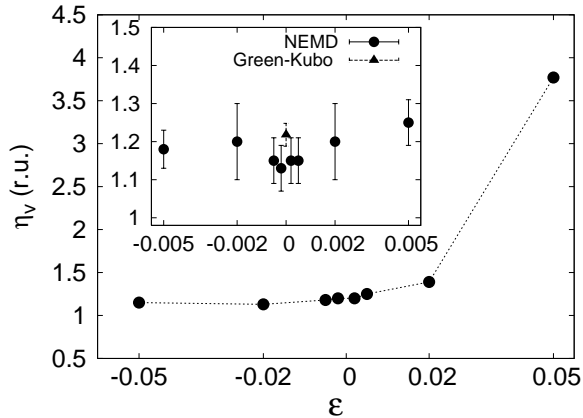}
\caption{Values of ${\eta}_v$ calculated in the present work. In the inset we present a magnification of the small ${\epsilon}$ region. A good agreement with the Green-Kubo result is obtained.}
\label{graf-finale}
\end{center}
\end{figure}
Finally, in table \ref{tab-finale} and in Fig.\ref{graf-finale} we report the values of ${\eta}_v(\epsilon)=\lim_{t\to\infty}\eta_v(t,\epsilon)$. As expected the data in the small $\epsilon$ region ($|\epsilon|\leq0.005$) are in agreement with the Green-Kubo estimate of the viscosity. Although equilibrium and NEMD data are in agreement within error bars, ${\eta}_v(\epsilon)$ data for $|\epsilon|\leq0.005$ exhibit an unexpected small error bar and appear to be systematically below the Green Kubo prediction. This is probably a small bias of our extrapolation procedure. The large noise in the tail of the response function enforces us to perform the fit in a time interval considerably smaller than for larger perturbations (see table $\ref{tab-finale}$). This might lead to underestimated errors (see Fig.\ref{graf-all-Int}) and to an estimate of the asymptotic value slightly lower than the correct value.
\begin{table}
\caption{\label{tab-finale}In the second and in the third column we report the data for $\Delta(0)$ and $\Delta^{\infty}$, respectively (see Eqs.(\ref{Deltat}) and (\ref{DeltaInf})) while, in the last column, the bulk viscosity for the corresponding intensity of the perturbation is shown. The time interval used for fitting the exponential function to the data is also reported in the second last column.}

% TABELLA ORIGINALE: SENZA i dati R(0) e R^{\infty}
%\begin{ruledtabular}
%\begin{center}
%\begin{tabular}{l l l}
%${\epsilon}$ & range for the fit & $\eta_v$  \\
%\hline
%0.05  & 0.44:2.66(500 pts)  & 3.77(2) \\
%0.02   & 0.44:2.66(500 pts) & 1.39(2) \\
%0.005  & 0.44:2.66(500 pts) & 1.25(6) \\
%0.002  & 0.44:2.22(400 pts) & 1.2(1)   \\
%0.0005 & 0.22:1.22(230 pts) & 1.15(6) \\
%0.0002 & 0.22:1.22(230 pts) & 1.15(6)  \\
%0(GK)  & ---                & 1.22(3)  \\
%-0.0002 &0.22:1.22(230 pts) & 1.13(6)  \\
%-0.0005 & 0.22:1.22(230 pts)& 1.15(6)  \\
%-0.002 & 0.44:2.22(400 pts) & 1.2(1) \\
%-0.005 & 0.44:2.66(500 pts) & 1.18(5) \\
%-0.02 & 0.44:2.66(500 pts)  & 1.13(2)  \\
%-0.05 & 0.44:2.66(500 pts)  & 1.15(2)  \\
%\end{tabular}
%\end{center}
%\end{ruledtabular}
%\end{table}

% Tabella date Eta + i dati R(0) e R^{\infty} presi dalla Ex tabella IV
\begin{ruledtabular}
\begin{center}
\begin{tabular}{l l l l l}
${\epsilon}$ & ${\Delta}(0)$ & ${\Delta}^{\infty}$ & range for the fit & $\eta_v$  \\
\hline
0.05  &  25.132(5)   & 13.757(6) & 0.44:2.66(500 pts)  & 3.77(2) \\
0.02  & 32.884(5)   & 21.382(7) & 0.44:2.66(500 pts) & 1.39(2) \\
0.005 &  37.778(4) & 25.61(2) & 0.44:2.66(500 pts) & 1.25(6) \\
0.002   &  38.863(4)   & 26.53(4)& 0.44:2.22(400 pts) & 1.2(1)   \\
0.0005  &  39.415(4)   & 27.1(2)& 0.22:1.22(230 pts) & 1.15(6) \\
0.0002  &  39.527(4)   &  26.6(4)& 0.22:1.22(230 pts) & 1.15(6)  \\
0(GK)   & --- & ---   & ---                & 1.22(3)  \\
-0.0002 & 39.663(5)  & 26.9(5)&0.22:1.22(230 pts) & 1.13(6)  \\
-0.0005 & 39.787(4)   & 27.2(2)& 0.22:1.22(230 pts)& 1.15(6)  \\
-0.002 & 40.351(4)  & 27.80(4)& 0.44:2.22(400 pts) & 1.2(1) \\
-0.005  & 41.499(5) & 28.74(2)& 0.44:2.66(500 pts) & 1.18(5) \\
-0.02 & 47.888(6)  & 34.116(7)& 0.44:2.66(500 pts)  & 1.13(2)  \\
-0.05 &  64.387(8) & 47.770(5)& 0.44:2.66(500 pts)  & 1.15(2)  \\
\end{tabular}
\end{center}
\end{ruledtabular}
\end{table}

%\begin{figure}[htbp]
%\begin{center}\includegraphics[width= 8 cm]{graf_Cort_all.eps}
%\caption{Dynamical response function $R(t)$ at several values of $\epsilon$. We report also the Green Kubo integrand $R^{GK}(t)$. In the limit of small $\epsilon$ the NEMD response approaches the Green Kubo behaviour.}
%\label{graf-Cot-all}
%\end{center}
%\end{figure}

\subsection{Non linear regime}\label{subsec:interpretation}
\begin{figure}[htbp]
\begin{center}
\includegraphics[width= 8 cm]{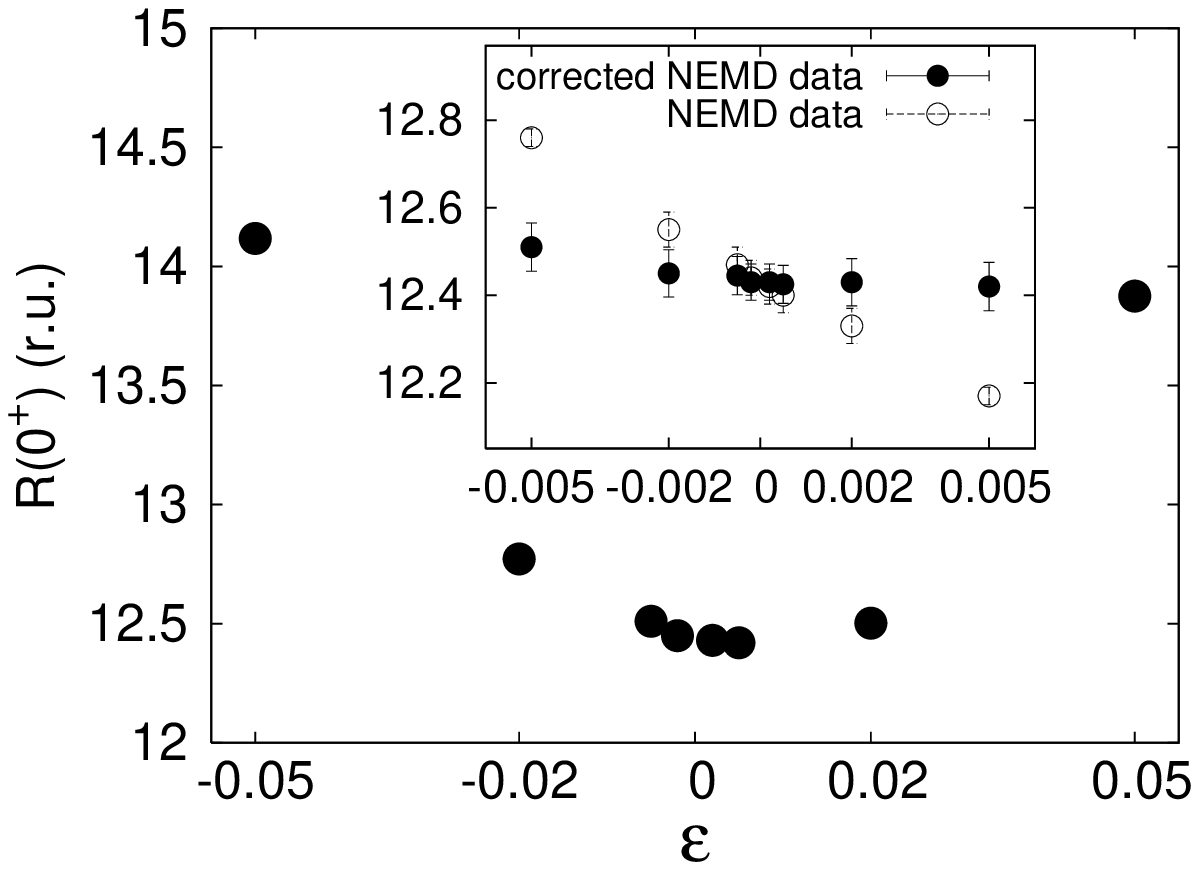}
\caption{Initial time response $R(0^+)$.The NEMD data are corrected by the thermodynamic term $3V\left(\frac{\partial R(0^+)}{\partial V}-p\frac{\partial R(0^+)}{\partial U}\right)_{\epsilon=0} \epsilon$, (see Eq.\ref{eq:R(t)_expansion}). In the inset we show an enlargement of the small ${\epsilon}$ region.}
\label{graf_Co-epsPro}
\includegraphics[width= 8 cm]{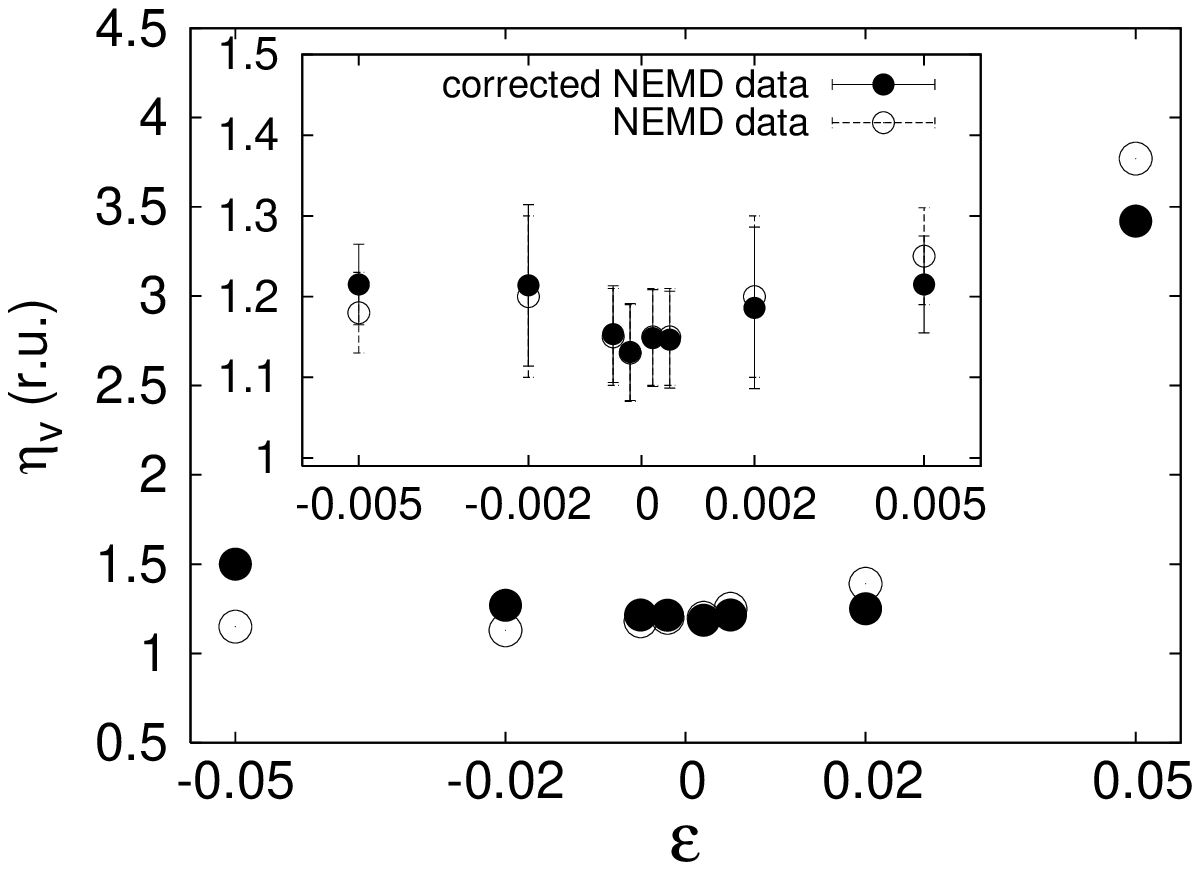}
\caption{Values of ${\eta}_v^{\epsilon}$ versus the perturbation. If the NEMD data are corrected by the thermodynamic term $3V\left(\frac{\partial \eta_v}{\partial V}-p\frac{\partial \eta_v}{\partial U}\right)_{\epsilon=0} \epsilon$, (see Eq. \ref{eq:eta_expansion}), then a large linear regime appears. In the inset we show an enlargement of the small ${\epsilon}$ region. }
\label{graf_finalePro}
\end{center}
\end{figure}

The results of previous section show the consistency between Green-Kubo and NEMD estimates of the bulk viscosity coefficient. However at variance with others coefficients, like for instance the shear viscosity, where a linear regime over several orders of magnitude of the intensity of the external perturbation is observed (up to roughly 0.05)\cite{FerrarioCiccottiHolianRyckaert91}, in the present case the linear regime is apparently much reduced. This can be clearly seen in figure \ref{graf-CoEpsilon} where the NEMD results for $R(0^+,\epsilon)$ matches the GK value with a finite slope suggesting that a linear expansion of the response function with the perturbation field is never justified. 
The same effect is seen for the viscosity in figure \ref{graf-finale} (see the inset), although it is less pronounced and the much larger noise makes the observation less conclusive.
In order to resolve this apparent paradox we must consider that the present perturbation, a contraction/expansion of the volume, excites the correct response and, at the same time, changes the thermodynamic state of the system. To correctly discuss the rheological non linear behavior of the fluid we should remove the pure thermodynamic contribution to the response. Let us consider the viscosity as a function of the perturbation, the internal energy and the volume of the system: $\eta_v=\eta_v(\epsilon,e,V)$. In the present case of impulsive perturbation we have $V(0^+)=V_0[1+3\epsilon+O(\epsilon^2)]$, and $E(0^+)=E_0-p_0[V(0^+)-V_0]+O(\epsilon^2)=E_0-3\epsilon p_0 V_0+O(\epsilon^2)$ where $E_0, V_0$, and $p_0$ represent the energy, volume and pressure of the equilibrium system respectively. The correct small $\epsilon$ expansion for the viscosity is therefore
\begin{equation}
\eta_v(\epsilon,e,V) =\eta_v + \left[\frac{\partial \eta_v}{\partial \epsilon}+3V\left(\frac{\partial \eta_v}{\partial V}-p\frac{\partial \eta_v}{\partial E}\right)\right]_{\epsilon=0} \epsilon+O(\epsilon^2) 
\label{eq:eta_expansion}
\end{equation}
A similar expansion holds for $R(t)$, in particular for its value at $t=0^+$
\begin{equation}
R(0^+;\epsilon,e,V) = R(0^+)_{\epsilon=0}+ \left[\frac{\partial R(0^+)}{\partial \epsilon}+3V\left(\frac{\partial R(0^+)}{\partial V}-p\frac{\partial R(0^+)}{\partial E}\right)\right]_{\epsilon=0} \epsilon+O(\epsilon^2) 
\label{eq:R(t)_expansion}
\end{equation}
We have performed a series of EMD simulations at volumes and internal energies around the thermodynamic point studied and we have estimated the derivatives in equations (\ref{eq:eta_expansion}), (\ref{eq:R(t)_expansion}) by the central difference formula. In table \ref{tab-derivate} we report the estimated values of the derivatives.
\begin{table}
\caption{\label{tab-derivate}Estimates of the thermodynamic derivatives in eqs. (\ref{eq:eta_expansion}) and (\ref{eq:R(t)_expansion}) at the considered state point as obtained by the central difference formula.}
\begin{ruledtabular}
\begin{center}
\begin{tabular}{c c}
$\frac{\partial \eta_v}{\partial V} = -0.0006(10) $ & $\frac{\partial \eta_v}{\partial E} =-0.0024(8) $ \\
$\frac{\partial R(0)}{\partial V} = -0.014(2)$ & $\frac{\partial R(0) }{\partial E} = 0.0021(9)$ 
\end{tabular}
\end{center}
\end{ruledtabular}
\end{table}
With those values the term in square brackets in eq. (\ref{eq:R(t)_expansion}) amounts to $-50\pm7$ to be compared with $-59\pm2$, the value of the estimated slope of the response in NEMD data (see figure \ref{graf_Co-epsPro}). As for the viscosity itself, the value in the square brackets in eq (\ref{eq:eta_expansion}) is $7\pm2$ to be compared with $6\pm1$ , the estimated slope of the viscosity in figure \ref{graf-finale}.
%\begin{figure}[htbp]
%\begin{center}
%\includegraphics[width= 8 cm]{graf_C0-E.eps}
%\caption{$ \left[3V \left(\frac{\partial R^{GK}(0)}{\partial V}  -p \frac{\partial R^{GK}(0)}{\partial U}\right)\right]$ = -52(3)(all data) ; = -50(7)(5 pt.s) From NEMD data =-59(2)}
%\label{grafCE}
%\includegraphics[width= 8 cm]{graf_C0-V.eps}
%\caption{}
%\label{grafCV}
%\end{center}
%\end{figure}
%\begin{figure}[htbp]
%\begin{center}
%\includegraphics[width= 8 cm]{graf_Eta-E.eps}
%\caption{$\left[3V \left(\frac{\partial \eta_v}{\partial V}  -p \frac{\partial \eta_v}{\partial U}\right)\right]$ = 3.44(all data); 7(5pt.s) }
%\label{grafEtaE}
%\includegraphics[width= 8 cm]{graf_Eta-V.eps}
%\caption{}
%\label{grafEtaV}
%\end{center}
%\end{figure}
When data for $R(0^+)$ and $\eta_v$ are corrected by these thermodynamic terms a large linear regime appears as reported in figures \ref{graf_Co-epsPro} and \ref{graf_finalePro}. This behaviour confirms the simple LJ fluid as a true linear fluid in a large range of perturbations, the genuine rheological non linear behaviour appearing only beyond $|\epsilon|\sim 0.02$.
%\begin{figure}[htbp]
%\begin{center}
%\includegraphics[width= 8 cm]{graf_Co-epsilonPROalldata.eps}
%\caption{Obtained using all data fit results}
%\label{graf_Co-epsPro_all}
%\includegraphics[width= 8 cm]{graf_finalePROalldata.eps}
%\caption{Obtained using all data fit results}
%\label{graf_finalePro_all}
%\end{center}
%\end{figure}

\section{Conclusions}\label{sec:conclusions}
In the present paper we have reported Non-Equilibrium Molecular Dynamics calculations of the bulk viscosity of the Lennard-Jones fluid at triple point. Among the transport coefficients of the simple fluid, the bulk viscosity was the only one for which NEMD results, from two independent previous studies, did not agree with the Green-Kubo estimates (see figure \ref{graf_eta_N}). Surprisingly, this unexpected failure of the Linear Response Theory remained unexplored for almost 25 years. In the present work we have resolved this apparent contradiction and found a full agreement between the NEMD and EMD estimates for the bulk viscosity. 

We have applied the Doll's perturbation field to excite the relevant flux in the system. At variance with the shear or elongational viscosity cases where the external perturbation is a superposition of a rotation and a deformation of the system at constant volume\cite{CiccottiPierleoniRyckaert90,PierleoniRyckaert91}, the field needed to excite the flux related to the bulk viscosity coefficient is a compression/expansion of the volume at constant shape. Such a perturbation excites the desired flux but also changes the thermodynamic state of the system. As a consequence, the relevant flux depends on the conditions at which the non equilibrium experiment is conducted and, similarly, the Green-Kubo formula depends on the equilibrium ensemble used\cite{McQuarry}. At equilibrium the microcanonical ensemble should be chosen to simplify the Green-Kubo analysis. As for the non equilibrium experiments, two different techniques can be applied. One can apply the \textit{stationary} NEMD method in which the system is driver toward a stationary non equilibrium state by applying a periodic compression/expansion of the system. If the heat produced by the external field is removed by a thermostatting mechanism, the steady state can be maintained in time and the bulk viscosity can be estimated as the time average of the relevant flux divided by the perturbation strength. The other possible route is the \textit{dynamical} NEMD in which a set of statistically uncorrelated replicas of the equilibrium system are subjected for $t\geq 0$ to the perturbation field and the evolution of the ensemble can be followed in time under the perturbed dynamics. The dynamical method is superior to the stationary method because not only steady state informations can be obtained but also the transient behavior can be fully characterized. Moreover within the \textit{dynamical} NEMD, the \textit{subtraction technique} can be used to perform the zero field strength limit and to extract the value of the linear transport coefficient. Another advantage of the dynamical method is that one can apply an impulsive perturbation rather than a periodic one. 
%In this case the transport coefficient is obtained integrating in time the relaxation of the relevant flux. 
Since the system is perturbed for a very short period of time (one step of our discrete dynamics) we do not need to introduce a thermostatting mechanism to perform a meaningful experiment.
%
%\textbf{Si potrebbe ribadire che la perturbazione sinusoidale, per essere implementata, necessita della conoscenza dell' equazione di stato $P_{eq}$ per valutare istante dopo istante $\bar p(\vec r,t)=P_{eq}(mn(\vec r,t),e(\vec r,t))$. Oppure, si puo` usare lo sviluppo di Peq() intorno ai valori di equilibrio ma conoscendo le sue derivate: in definitiva sarebbe molto complicato.}
%
Using dynamical NEMD with the subtraction technique and an impulsive form of the perturbation, we were able to explore a large range of perturbation strength and carefully study the small field regime. We have found a perfect agreement between the NEMD and the Green-Kubo estimates of the bulk viscosity. These estimates are also in agreement with previous values obtained by the Green-Kubo formula for various system sizes, while they do not agree with previous NEMD studies conducted by the stationary NEMD method.
Finally, by removing the thermodynamic contribution to the viscosity, we show that the Lennard-Jones fluid at triple point exhibits a large linear regime, in agreement with the results for the shear viscosity\cite{Ryckaertatall}.



%%%%%%%%%%%%%%%%%%%%%%%%%%%%%%%%%%%%%%%%%%%%%%%%%%%%%%%%%%%%%%%%%%%%
%%%%%%%%%%%%%%%%%%%%%%%%%%%%%%%%%%%%%%%%%%%%%%%%%%%%%%%%%%%%%%%%%%%%
\newpage
%\begin{widetext}
\appendix*
\section{Perturbation associated to viscous flux: ``Doll's'' }
In this section we will compute the response in the velocity field to the ``Doll's'' perturbation. By a straightforward application of the linear response theory we will find the result in Eq. (\ref{vVSu}). It proves that the ``Doll's'' perturbation is the perturbation producing every kind of viscous flux.

Let us consider a system of $N$ particles of mass $m$ with Hamiltonian of such a system will be:  
\begin{equation}
\hat H^{(0)} = \sum_{i=1}^{N}\frac{\dot{\vec p}_i^2}{2m} + \sum_{j \neq i}^N \phi(\mid\vec r_{ij}\mid) \label{hamilton0}
\end{equation}
where $\vec r_{ij}=\vec r_i-\vec r_j$ and $\phi(r)$ is the pair potential. We add a perturbation term of the form: 
\begin{equation}
\hat H^{(I)}(\Gamma,t)=\int_V \underline{\underline \alpha}(\vec r):{\underline{\underline \Phi}}(\vec r,t) d\vec r
 \label{hamiltonI}
\end{equation}
where  $V$ is the volume of the system, $\underline{\underline \alpha}(\vec r)$ is a local dynamical variable and ${\underline{\underline \Phi}}(\vec r,t)$ is the external field, dependent on time $t$ and space $\vec r$. We assume that ${\underline{\underline \Phi}}$ is proportional to a small parameter $\epsilon$ defining the magnitude of the perturbation. 
The linear response theory states that the effect of $\hat H^{(I)}$ on a given observable $\hat O$ of the system is:
\begin{equation}
O(\vec r,t) = <\hat O>_0  - \beta \int_0^{\infty} d\tau \int_V d\vec s \;<\hat O (\vec 0,0)\dot{\underline{\underline \alpha}}(\vec{s},\tau ) \rho_0>:{\underline{\underline\Phi}}(\vec{r}-\vec{s},t-\tau) + {\mathcal O} (\epsilon^2)  \label{RispLin}
\end{equation}
In the present case $\hat H^{(I)}$ is given by Eq.\ref{dolls}, therefore we set:
$ \underline{\underline \alpha}=\sum_i^N \vec r_i \vec p_i \delta(\vec r_i - \vec r )$ and ${\underline{\underline \Phi}}=\left(\vec{\nabla}\vec u(\vec r,t)\right)^T$.

We have to calculate the velocity field $\vec v$ for the system subjected to the perturbation. By following Irving-Kirkwood theory, we can identify this field by exploiting the equation:
$$
\vec J^m(\vec r,t)= m(\vec r,t)\vec v(\vec r,t)= <\vec{g}(\vec r)>_t
$$
where we have introduced the flux of momentum $\vec J^m$ and the corresponding dynamical variable $\vec{g}$:
\begin{equation}
\vec{g}(\vec r) = \sum _i^N m   \dot{\vec r}_i \delta(\vec r_i - \vec r ) \label{gdef}
\end{equation}
Expanding the last expression in series of the small parameter $\epsilon$, we find:
$$ 
\vec v(\vec r,t)= \frac{\vec J^m(\vec r,t)} {m(\vec r,t)} =\frac{ \vec J_m^{(I)}(\vec r,t) + {\mathcal O}(\epsilon^2) }{ m^{(0)}(\vec r,t)+ m^{(I)}(\vec r,t)+ {\mathcal O} (\epsilon^2) }=
$$
$$
 =\frac{V}{mN}\vec J_m^{(I)}(\vec r,t) + {\mathcal O} (\epsilon^2)
$$
where we have assumed that in the limit $\epsilon \rightarrow 0$: $\vec J^m(\vec r,t) \rightarrow 0$ and $m(\vec r,t) \rightarrow \frac{mN}{V}$.
Now, by applying the linear response theory Eq. (\ref{RispLin}), we find the following equation for the $\alpha$ component of the field $\vec v$:
$$
v_\alpha(\vec r,t) =  - \frac{V}{mN}  \beta \int_0^{\infty} d\tau \int_V d\vec s \;< g_\alpha (\vec 0,0) \dot{\alpha}_{\mu \nu}(\vec{s},\tau )>_0 {\nabla}_\mu u_\nu(\vec{r}+\vec{s},t-\tau) + {\mathcal O} (\epsilon^2)
$$
Finally, by performing an integration by parts we obtain: 
$$
v_\alpha(\vec r,t) =  \frac{V}{mN}  \beta \int_0^{\infty} d\tau \int_V d\vec s \;< g_\alpha (\vec 0,0) {\nabla}_\mu  \dot{\alpha}_{\mu \nu}(\vec{s},\tau ) >_0  u_\nu(\vec{r}-\vec{s},t-\tau) + {\mathcal O} (\epsilon^2)
$$
%ovvero:
%$$
%\vec v(\vec r,t) =  \frac{V}{mN}  \beta \int_0^{\infty} d\tau \int_V d\vec s \;<\vec{\hat g}(\vec 0,0) \vec{\nabla} \cdot \dot{\hat{\underline{\underline \alpha}}}(\vec{s},\tau ) >_0  \cdot \vec u(\vec{r}+\vec{s},t-\tau) + {\mathcal O} (\epsilon^2)
%$$
which is a convolution in space and time. Its Fourier transform is
\begin{equation}
{\tilde v_{\alpha}}(\vec k,\omega)={\sigma}_{\alpha \beta}(\vec{k},\omega) \tilde{u}_{\beta}(\vec k,\omega) \label{RLk_su_v}
\end{equation}
where 
$$
{\sigma}_{\alpha\beta}(\vec{k},\omega)=\frac{V}{mN} \beta \int_{0}^{+\infty}d\tau \int d\vec s \;e^{i(-\vec k \cdot \vec s+\omega \tau) } <g_{\alpha} (\vec 0,0) {\nabla}_{\mu} \dot{\alpha}_{\mu \beta}(\vec{s},\tau ) >_0=
$$
$$
=-\frac{V}{mN} \beta \int_{0}^{+\infty}d\tau  \;e^{i\omega \tau} <g_{\alpha} (\vec 0,0) (\tilde{ \nabla_{\mu}\dot{\alpha} }_{\mu\beta})(\vec{k},\tau ) >_0
$$
%$$
%=-\frac{V}{mN} \beta \int_{0}^{+\infty}d\tau  \;e^{i\omega \tau} <{\hat g}_{\alpha} (\vec 0,0)  i\;k_{\mu} \dot{\hat{\alpha}}_{\mu\beta}(\vec{-k},\tau ) >_0
%$$
where we have performed the integral in $d\vec s$ in the second last equality.

In order to evaluate ${\sigma}_{\alpha\beta}(\vec{k},\omega)$, we have to calculate the Fourier transform $(\tilde{ \nabla_{\mu}\dot{\alpha} }_{\mu\beta})$. In the following equation, we report the result of this calculation that will be demonstrated at the end of this paragraph. In the case with $k<<a$,  where ``$a$'' is the mean free path, we will find: 
\begin{equation}
(\tilde{ {\nabla}_{\mu} \dot{\alpha}}_{\mu\beta} )(\vec{k},\tau )=- \dot{{\tilde g}}_{\beta}(\vec k,t) -  i k_{\mu} \cdot \sum_i^N (i\vec{k} \cdot \dot{\vec r}_i) ( \vec r_i \vec p_i )_{\mu\beta} \;e^{-i\vec k \cdot \vec r_i} \label{LastDimEq}
 \end{equation}
With the aid of the last equation, we can determine ${\sigma}_{\alpha\beta}$, defined in Eq. \ref{RLk_su_v}, in the limit $k \rightarrow 0$:
\begin{eqnarray}
\lim_{k \rightarrow 0}{\sigma}_{\alpha\beta}(\vec{k},\omega) & = & \frac{V}{mN} \beta \int_{0}^{+\infty}d\tau \;e^{i \omega \tau } <g_{\alpha} (\vec 0,0) \dot{\tilde {g}}_{\beta}(\vec{k}=0,\tau ) >_0= \nonumber \\
& = & -\frac{V}{mN} \beta \int_{0}^{+\infty}d\tau \;i \omega e^{i \omega \tau } < g_{\alpha} (\vec 0,0) {\tilde{g}}_{\beta}(\vec{k}=0,\tau ) >_0 \label{sigmaaux}
\end{eqnarray}
%\textbf{Come \`e scomparsa la condizione iniziale visto che stiamo usando (Fourier)-Laplace?}\\
The ensemble average $<g_{\alpha} (\vec 0,0) \tilde{g}_{\beta}(\vec{k}=0,\tau )>_0$ can be calculated as follows: 
$$
<g_{\alpha}(\vec r,0) \tilde{g}_{\beta}(\vec{k}=0,\tau )>_0=<g_{\alpha} (\vec r,0)\int d \vec {s} g_{\beta}(\vec s,\tau )>_0=
$$
$$
=< \sum _i^N m   \dot{r}_{i\alpha} \delta(\vec r_i - \vec r ) \sum _j^N m   \dot{r}_{j\beta}(\tau)>_0
$$
where we have used the definition of $\vec{g}$ given by Eq.\ref{gdef}. Now, we note that the quantity $\sum _j^N m  \dot{r}_{j\beta}(\tau)$ corresponds to the total momentum of the system, therefore it is independent of the time. On the other hand, the overall average on the equilibrium ensemble have to be $\vec r$ independent as well. Therefore the following relation holds:
$$
<g_{\alpha}(\vec r,0) \tilde{g}_{\beta}(\vec{k}=0,\tau )>_0=\frac{1}{V} \int_V d\vec r\;< \sum _i^N m   \dot{r}_{i\alpha} \delta(\vec r_i - \vec r ) \sum _j^N m   \dot{r}_{j\beta}>_0
$$
$$
=\frac{1}{V} < \sum _i^N m   \dot{r}_{i\alpha}  \sum _j^N m   \dot{r}_{j\beta}>_0=\frac{m}{V} < \sum _i^N m   \dot{r}_{i\alpha}^2>_0\delta_{\alpha \beta} = m \frac{N}{\beta V}\delta_{\alpha \beta}
$$
In the last equality we have also applied the equipartition theorem $<m\dot{r}_{i\alpha}^2>_0=K_bT=\frac{1}{\beta}$. By replacing the last result in Eq. (\ref{sigmaaux}) we finally find:
$$
\lim_{k \rightarrow 0} \underline{\underline \sigma}(\vec{k},\omega)= - \int_{0}^{+\infty}d\tau \;i \omega e^{i \omega \tau }{\underline{\underline I}} = -i \omega \int_{-\infty}^{+\infty}d\tau \; e^{i \omega \tau }\theta(\tau){\underline{\underline I}}=
$$
$$
=-i \omega \tilde{\theta}{\underline{\underline I}}= \tilde{\left(\frac{d\theta}{dt} \right)}{\underline{\underline I}}=\tilde \delta{\underline{\underline I}}={\underline{\underline I}}\label{sigmaaux2}
$$
where $\theta$ stands for the step function.
By comparing this relation to Eq. (\ref{RLk_su_v}) we obtain the final result $\tilde {\vec v}(\vec 0,\omega) = \tilde{\vec u}(\vec 0,\omega) \label{v-vs-u}$ that proves the correspondence between the field $\vec u$ involved in the ``Doll's'' perturbation and the macroscopic velocity field $\vec v$ induced in the system by the perturbation.
%%%%%%%%%%%%%%%%%%%%%%% alfa punto %%%%%%%%%%%%%%%%%%%%%%%%

In order to complete the paragraph we have to demonstrate equation Eq. (\ref{LastDimEq}). First of all, we calculate  the derivative $\dot{\underline{\underline \alpha}}$ as follows:
$$
\dot{\underline{\underline \alpha}}(\vec{r},t)=\sum_i^N(\dot{\vec r}_i \vec p_i +\vec r_i \dot{\vec p}_i)\delta(\vec r_i - \vec r ) + \vec r_i \vec p_i\left( \dot{\vec r}_i  \cdot  \frac{\partial  }{\partial \vec r_i}\delta(\vec r_i - \vec r )  \right) =
$$
$$
= \sum_i^N  \left(m \dot{\vec r}_i \dot{\vec r}_i - \vec r_i \sum_{j\neq i}^N\frac{\partial \phi(\mid\vec r_{ij}\mid)}{\partial \vec r_i} - \vec r_i \vec p_i (\dot{\vec r}_i\cdot\vec{\nabla}) \right) \delta(\vec r_i - \vec r ) %+  {\mathcal O} (\epsilon^2)
$$
where we have applied the relation $\frac{\partial  }{\partial \vec r_i} \delta(\vec r_i - \vec r )= -\vec{\nabla} \delta(\vec r_i - \vec r )$.

Then, we calculate the Fourier transform on the space variable $\vec r$:
\begin{equation}
\tilde{\dot{\underline{\underline \alpha}}}(\vec{k},t)=  \sum_i^N \left( m \dot{\vec r}_i \dot{\vec r}_i - \vec r_i \sum_{j\neq i}^N\frac{\partial \phi(\mid\vec r_{ij}\mid)}{\partial \vec r_i} +i \vec r_i \vec p_i (\dot{\vec r}_i \cdot \vec k) \right)e^{i\vec k \cdot \vec r_i} \label{tildedotalfa} 
\end{equation}
By focusing on the former equation term that involves the derivative of the potential, we find:
$$
\sum_i^N\sum_{j\neq i}^N \vec r_i \frac{\partial \phi(\mid\vec r_{ij}\mid)}{\partial \vec r_i}e^{i\vec k \cdot \vec r_i}=
$$
$$
=\frac{1}{2}\sum_i^N\sum_{j\neq i}^N\left(\vec r_i \frac{\partial \phi(\mid\vec r_{ij}\mid)}{\partial \vec r_i}e^{i\vec k \cdot \vec r_i} + \vec r_j \frac{\partial \phi(\mid\vec r_{ji}\mid)}{\partial \vec r_j}e^{i\vec k \cdot \vec r_j}   \right)=
$$
$$
=\frac{1}{2}\sum_i^N\sum_{j\neq i}^N\left(\vec r_i e^{i\vec k \cdot \vec r_i} - \vec r_j e^{i\vec k \cdot \vec r_j}  \right)\frac{\partial \phi(\mid\vec r_{ij}\mid)}{\partial \vec r_i}=
$$
$$
=\frac{1}{2}\sum_i^N\sum_{j\neq i}^N\;e^{i\vec k \cdot \vec r_j}\left(\vec r_i e^{i\vec k \cdot \vec r_{ij}} - \vec r_j  \right)\frac{\partial \phi(\mid\vec r_{ij}\mid)}{\partial \vec r_i}=
$$
$$
=\frac{1}{2}\sum_i^N\sum_{j\neq i}^N\;e^{i\vec k \cdot \vec r_j}\left[\vec r_{ij}  + \vec r_i (i\vec k \cdot \vec r_{ij} + {\mathcal O}({\vec k \cdot \vec r_{ij}}^2) )\right]\frac{\partial \phi(\mid\vec r_{ij}\mid)}{\partial \vec r_i}
$$
By means of this result and under the hypothesis of small wave vector limit $ka<<1$, Eq. (\ref{tildedotalfa}) can be written as follows:
\begin{equation}
\tilde{\dot{\underline{\underline \alpha}}}(\vec{k},t) = 
 \sum_i^N  m \dot{\vec r}_i \dot{\vec r}_i \;e^{i\vec k \cdot \vec r_i} - \frac{1}{2}\sum_{j\neq i}^N\;e^{i\vec k \cdot \vec r_j}\vec r_{ij} \frac{\partial \phi(\mid\vec r_{ij}\mid)}{\partial \vec r_i}        +  \sum_i^N (i\vec k\cdot \dot{\vec r}_i) ( \vec r_i \vec p_i ) \;e^{i\vec k \cdot \vec r_i} \label{tildedotalfa2} 
\end{equation}
Eq. (\ref{tildedotalfa2}) implies that the Fourier transform of the gradient of $\dot{\underline{\underline \alpha}}$ fulfills the following equation:
$$
( \tilde{ \vec{\nabla}\cdot\dot{\underline{\underline \alpha}} }\;  ) (\vec{k},\tau )= \left[- i \vec k \cdot \left(\sum_i^N  m \dot{\vec r}_i \dot{\vec r}_i \;e^{i\vec k \cdot \vec r_i} - \frac{1}{2}\sum_{j\neq i}^N\;e^{i\vec k \cdot \vec r_j}\vec r_{ij} \frac{\partial \phi(\mid\vec r_{ij}\mid)}{\partial \vec r_i}\right) \right]-
$$
$$ - i \vec k \cdot\sum_i^N (i\vec k\cdot \dot{\vec r}_i) ( \vec r_i \vec p_i ) \;e^{i\vec k \cdot \vec r_i}
$$
This result correspond to the required Eq. (\ref{LastDimEq}) providing that the quantity in square brackets is equal to  $-\tilde{\dot{\vec g}}(\vec k,t)$. This can easily verified as follows:
$$
\tilde{\dot{\vec g}}(\vec k,t)= \frac{d}{dt}\sum_{i=1}^N m \dot{\vec r}_i e^{i\vec k \cdot \vec r_i}=\sum_{i=1}^N \left(  m \dot{\vec r}_i (i\vec k \cdot \dot{\vec r}_i) + m  \ddot{\vec r}_i  \right)  e^{i\vec k \cdot \vec r_i}=
$$
$$
=\sum_{i=1}^N \left( m \dot{\vec r}_i (i\vec k \cdot \dot{\vec r}_i)  -\sum_{j\neq i}^N\frac{\partial \phi(\mid\vec r_{ij}\mid)}{\partial \vec r_i}   \right)  e^{i\vec k \cdot \vec r_i}=
$$
$$
=\sum_{i=1}^N i\vec k \cdot  \left( m \dot{\vec r}_i \dot{\vec r}_i\right) e^{i\vec k \cdot \vec r_i} - \frac{1}{2}\sum_{i=1}^N \sum_{j\neq i}^N \left( \frac{\partial \phi(\mid\vec r_{ij}\mid)}{\partial \vec r_i} e^{i\vec k \cdot \vec r_i} +\frac{\partial \phi(\mid\vec r_{ji}\mid)}{\partial \vec r_j} e^{i\vec k \cdot \vec r_j} \right)=
$$
$$
=\sum_{i=1}^N i\vec k \cdot  \left( m \dot{\vec r}_i \dot{\vec r}_i\right) e^{i\vec k \cdot \vec r_i} - \frac{1}{2} \sum_{i=1}^N \sum_{j\neq i}^N \frac{\partial \phi(\mid\vec r_{ij}\mid)}{\partial \vec r_i} \left( e^{i\vec k \cdot \vec r_i} - e^{i\vec k \cdot \vec r_j} \right)=
$$
$$
=\sum_{i=1}^N i\vec k \cdot  \left( m \dot{\vec r}_i \dot{\vec r}_i\right) e^{i\vec k \cdot \vec r_i} - \frac{1}{2} \sum_{i=1}^N \sum_{j\neq i}^N \frac{\partial \phi(\mid\vec r_{ij}\mid)}{\partial \vec r_i} e^{i\vec k \cdot \vec r_j} \left(  e^{i\vec k \cdot \vec r_{ij}}-1 \right)=
$$
$$
=\sum_{i=1}^N i\vec k \cdot  \left( m \dot{\vec r}_i \dot{\vec r}_i\right) e^{i\vec k \cdot \vec r_i} - \frac{1}{2} \sum_{i=1}^N \sum_{j\neq i}^N \frac{\partial \phi(\mid\vec r_{ij}\mid)}{\partial \vec r_i} e^{i\vec k \cdot \vec r_j} \left(  i\vec k \cdot \vec r_{ij} + {\mathcal O}((i\vec k \cdot \vec r_{ij})^2 ) \right)
$$
In the limit $ka<<1$ we finally get
$$
\tilde{\dot{\vec g}}(\vec k,t)=i \vec k \cdot  \sum_{i=1}^N  \left( m (\dot{\vec r}_i \dot{\vec r}_i)\right) e^{i\vec k \cdot \vec r_i} + \frac{1}{2} \sum_{i=1}^N \sum_{j\neq i}^N \frac{\partial \phi(\mid\vec r_{ij}\mid)}{\partial \vec r_i} e^{i\vec k \cdot \vec r_j} \left(  i\vec k \cdot \vec r_{ij} \right) =
$$
$$
= \left[ i\vec k \cdot  \left(\sum_{i=1}^N \left( m \dot{\vec r}_i \dot{\vec r}_i\right) e^{i\vec k \cdot \vec r_i} + \frac{1}{2} \sum_{i=1}^N \sum_{j\neq i}^N  \vec r_{ij} \frac{\partial \phi(\mid\vec r_{ij}\mid)}{\partial \vec r_i} e^{i\vec k \cdot \vec r_j} \right)\right] 
$$

%\end{widetext}

%\begin{acknowledgments}

%\end{acknowledgments}


\begin{thebibliography}{100}


%\bibitem{Alder}
%B.J.Alder, D.M.Gass, T.E.Wainwright, J.Chem.Phys. \textbf{53}, p.3813 (1970).
\bibitem{CiccottiHoover}
G. Ciccotti and W.G. Hoover, \textit{Molecular Dynamics Simulation of Statistical-Mechanical Systems}, International School of Physics, Course XCVII, (North Holland, Amsterdam 1986).

%\bibitem{Astro}  
%K. Paech and S. Pratt, Phys.Rev.C \textbf{74}, 014901 (2006).

\bibitem{Green} 
M.S.Green, J.Chem.Phys.\textbf{19}, p.34 (1951).

\bibitem{Kubo}
%R Kubo,  Rep. Prog. Phys. \textbf{29}, p.255 (1966)
R Kubo,  J. Phys. Soc. Jap. \textbf{12}, p.570 (1957)

\bibitem{HansenMcDonald}
J. P. Hansen and I. R. Mc Donald: \emph{Theory of Simple Liquids}, Academic Press (3rd edition 2006)

\bibitem{Trozzi}  
 C. Trozzi and G. Ciccotti, Phys.Rev.A \textbf{29}, 925 (1984).

\bibitem{Gallico}  
 A. Tenenbaum, G. Ciccotti, and R. Gallico, Phys.Rev.A \textbf{25}, 2778 (1982).

\bibitem{Koplik88}
J. Koplik, J. R. Banavar and J. F. Willemsen, Phys. Rev. Letts. \textbf{60}, 1282 (1988).

\bibitem{EvansMorris}
D. J. Evans and G. P. Morriss: \emph{Statistical Mechanics of NonEquilibrium Liquids}, Academic Press, London (1990)

\bibitem{LeesEdwards} 
A.W. Lees e S.F. Edwards, J.Phys C \textbf{5}, p.1921 (1972).

\bibitem{He'84}  
D.Heyes, J.Chem.Soc.Far.Tr.\textbf{80},p.1363 (1984).

\bibitem{CiccottiJacucciPRIMO} 
G.Ciccotti e G.Jacucci, Phys. Rev. Lett. \textbf{35}, p.789 (1975).

\bibitem{CiccottiJacucciTERZO}
G.Ciccotti, G.Jacucci, McDonald, J. Stat. Phys. \textbf{21}, p.1 (1979).

\bibitem{MassobrioCiccotti84}
C. Massobrio and G. Ciccotti, Phys. Rev. A \textbf{30}, 3191 (1984).

\bibitem{HooverCiccottiPaoliniMassobrio85}
W.G.Hoover, G.Ciccotti, G.V.Paolini and C.Massobrio, Phys. Rev. A \textbf{32}, 3765 (1985).

\bibitem{PaoliniCiccotti87}
G.V.Paolini and G.Ciccotti, Phys.Rev. A \textbf{35},  5156 (1987).

\bibitem{PierleoniCiccottiBernu87}
C.Pierleoni, G.Ciccotti and B.Bernu, Europhys. Letts. \textbf{4}, 1115 (1987).

\bibitem{PierleoniCiccotti90}
C. Pierleoni and G. Ciccotti, J. Phys. Cond. Matt. \textbf{2}, 1315 (1990).

\bibitem{HonkonnouPierleoniRyckaert92}
M.N. Honkonnou, C. Pierleoni and J.P. Ryckaert, J. Chem. Phys. \textbf{97}, 9335 (1992).

\bibitem{RyckaertCiccotti89}
L.P. Ryckaert, A. Bellemans, G. Ciccotti, G.V. Paolini, Phys.Rev. A \textbf{39}, p.259 (1989) 

\bibitem{M'05}
K.Meier, A.Laesecke, S.Kabelac, J.Chem.Phys. \textbf{122}, 014513 (2005).

\bibitem{LVK'73} 
D.Levesque,L. Verlet,J.Kurkijarvi, Phys.Rev. A  \textbf{7}, p.1690 (1973).

\bibitem{LV'87} 
D.Levesque, L. Verlet, Mol.Phys.  \textbf{61}, p.143 (1987).

\bibitem{SH'85}
M.Schoen C.Hoheisel, Mol.Phys.\textbf{56}, p. 653 (1985).

\bibitem{Hohe'87}
C.Hoheisel, J.Chem.Phys.  \textbf{86}, p.2328 (1987).

\bibitem{HVS'87} 
C.Hoheisel, R.Vogelsang, M.Schoen, J.Chem.Phys.  \textbf{87}, p.7195 (1987).

\bibitem{Hoover80_2} 
B.Hoover, D.J.Evans,R.B.Hickman at al., Phys.Rev. A \textbf{22}, p.1690 (1980).

%%%%%%%%%%%%%%%%%%%%%%%%

%\bibitem{Malbrunot}
%P.Malbrunot, A. Boyer, E. Charles, H. Abachi, Phys.Rev. A \textbf{27}, p.1523 (1983).


%\bibitem{CiccottiJacucciSECONDO}
% G.Ciccotti, G.Jacucci, McDonald, Phys. Rev. A \textbf{13}, p.426 (1976).

%\bibitem{CiccottiJacucciTERZO}
%G.Ciccotti, G.Jacucci, McDonald, J. Stat. Phys. \textbf{21}, p.1 (1979).


%\bibitem{MAZURbook}
%S.R.de Groot e P.Mazur: \emph{Non-equilibrium thermodynamics}, Dover Publications, Inc.,New York (First edition 1961)

\bibitem{Kirk}
 Irving e Kirkwood, J.Chem.Phys \textbf{18}, p.6 (1950).

\bibitem{Mori}
H.Mori,  Progress of Theoretical Physics \textbf{28}, p.763 (1962).

\bibitem{McQuarry}
McQuarry, \emph{Statistical Mechanics}, Harper \& Row, New York (1976)

\bibitem{Luttinger}
J.M.Luttinger, Phys. Rev. \textbf{135},p.1505 (1964).

\bibitem{CiccottiPierleoniRyckaert90}
G. Ciccotti, C. Pierleoni and J.P. Ryckaert, "Theoretical Foundation and Rheological Application of Nonequilibrium Molecular Dynamics", in "Microscopic Simulation of Complex Hydrodynamic Phenomena". Mareschal M. and Holian B.L. eds., Plenum Press (1991).

\bibitem{Hoover80_1}
B. Hoover, A.Ladd, R.B.Hickman at al., Phys.Rev. A  \textbf{21}, p.1756 (1980)

\bibitem{ryckaert90Aachen}
J.P. Ryckaert, Ber. Bunsenges. Phys. Chem \textbf{94}, 256 (1990).

\bibitem{Ryckaertatall}
J.P. Ryckaert, A.Bellemans, G.Ciccotti and G.V. Paolini, Phys. Rev. Lett. \textbf{60}, 128 (1988). 

\bibitem{FerrarioCiccottiHolianRyckaert91}
M. Ferrario, G. Ciccotti, B. L. Holian and J.P. Ryckaert, Phys. Rev. A \textbf{44}, 6936 (1991).

\bibitem{PierleoniRyckaert91}
C. Pierleoni and J.P. Ryckaert, Phys. Rev. \textbf{A} 44, 5314 (1991).

\end{thebibliography}

\end{document}